\documentclass[final]{aipproc}
\layoutstyle{6x9}

\begin{document}

\title{Status of Dynamical Coupled-Channel analysis by Collaboration$@$EBAC}


\author{T.-S. H. Lee}{
address={Physics Division, Argonne National Laboratory, Argonne, 
Illinois 60439, USA}}

\begin{abstract}
The development and results of the Dynamical Coupled-Channels
analysis by a collaboration at
 the Excited Baryon Analysis Center (EBAC) are reported.
\end{abstract}

\maketitle

\section{Introduction}

In this contribution, we report on the development and results from a
dynamical coupled-channel (DCC) analysis by a collaboration associated with the
Excited Baryon Analysis Center (EBAC) at Jefferson Lab  during
the period January, 2006 - March, 2012.
This  project has three components :
(1) perform a  dynamical coupled-channels analysis 
on the {\it world} data on meson production reactions from the nucleon
to determine the meson-baryon partial-wave amplitudes, (2)
extract the $N^*$ parameters 
from the determined partial-wave amplitudes, and (3) 
investigate the interpretations of the extracted $N^*$ properties 
in terms of the available hadron models and Lattice QCD.

\section{DCC model}

The  DCC analysis by the collaboration$@$EBAC  is  based on 
an extension~\cite{msl07} 
of the dynamical
model for the $\Delta$(1232) resonance  developed \cite{sl96} by an
Argonne National Laboratory-Osaka University (ANL-Osaka) collaboration.
 Within the extended ANL-Osaka formulation~\cite{msl07}, 
the reaction amplitudes $T_{\alpha,\beta}(p,p';E)$ in each partial-wave are calculated from
the following coupled-channels integral equations,
\begin{eqnarray}
T_{\alpha,\beta}(p,p';E)&=& V_{\alpha,\beta}(p,p') + \sum_{\gamma}
 \int_{0}^{\infty} q^2 d q  V_{\alpha,\gamma}(p, q )
G_{\gamma}(q ,E)
T_{\gamma,\beta}( q  ,p',E) \,, \label{eq:cct}\\
V_{\alpha,\beta}&=& v_{\alpha,\beta}+
\sum_{N^*}\frac{\Gamma^{\dagger}_{N^*,\alpha} \Gamma_{N^*,\beta}}
{E-M^*} \,,
\label{eq:ccv}
\end{eqnarray}
where $\alpha,\beta,\gamma = \gamma N, \pi N, \eta N, KY, \omega N$, and
$\pi\pi N$ which has  $\pi \Delta, \rho N, \sigma N$ resonant components,
$v_{\alpha,\beta}$ are meson-exchange interactions deduced from
phenomenological Lagrangian, $\Gamma_{N^*,\beta}$  describes
the excitation of the nucleon to a bare $N^*$ state with a mass
$M^*$, and $G_{\gamma}(q ,E)$ is a meson-baryon propagator. 
The DCC model, defined by Eqs.~(\ref{eq:cct}) and~(\ref{eq:ccv}), 
satisfies two- and three-body unitarity conditions.


In order to determine the parameters associated with the strong-interaction
parts of  $V_{\alpha,\beta}$ in Eq.~(\ref{eq:ccv}), the considered
DCC model was first applied
to fit the $\pi N$ elastic scattering up to invariant mass $W = 2 $ GeV.
For simplicity, $KY$ and $\omega N$ channels were not included during
the development stage in 2006-2010.
The electromagnetic parts of  $V_{\alpha,\beta}$ were then determined by fitting
the data of $\gamma p \rightarrow \pi^0p, \pi^+n$ and $p(e,e'\pi^{0,+})N$.

The resulting 6-channel model was then tested by comparing 
the predicted $\pi N, \gamma N \rightarrow \pi\pi N$ production cross sections with the data.
In parallel to analyzing the data, a procedure
to analytically continue Eqs.~(\ref{eq:cct}) and~(\ref{eq:ccv}) 
to the complex energy plane was developed to extract
the positions and residues of nucleon resonances.

In the following, we present a sample of results from these efforts.

\subsection{Results for single pion production reactions}

In fitting the  $\pi N$ elastic scattering, we found that one or two bare $N^*$ states 
were needed in each partial wave. 
The coupling strengths of the $N^*\rightarrow MB$ vertex interactions
$\Gamma_{N^*,MB}$ with $MB=\pi N, \eta N, \pi\Delta, \rho N, \sigma N$
were then determined in the $\chi^2$-fits to the data.
Our results were given in Ref.~\cite{jlms07}.

\begin{figure}[t]
\includegraphics[clip,height=.35\textheight]{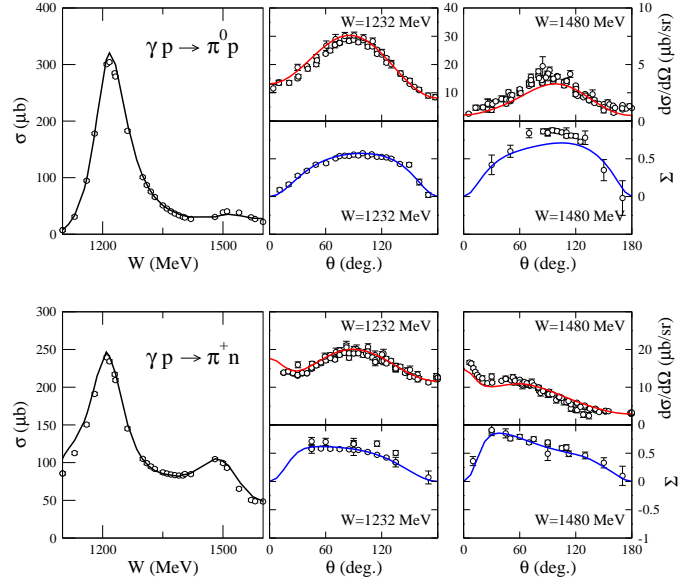}
\caption{The DCC results~\cite{jlmss08} of total cross sections ($\sigma$), 
differential cross sections ($d\sigma/d\Omega)$, and photon
asymmetry ($\Sigma$) of $\gamma p\rightarrow \pi^0p$ (upper parts), 
$\gamma p \rightarrow \pi^+n$ (lower parts).}
\label{fig:gnpin}
\end{figure}

Our next step was to determine the bare $\gamma N \rightarrow N^*$
interaction $\Gamma_{N^*,\gamma N}$ by fitting the $\gamma p \rightarrow
\pi^0p$ and $\gamma p \rightarrow \pi^+n$ data. 
We found~\cite{jlmss08} that we were able to fit the data only
up to invariant mass $W = 1.6$ GeV, mainly because we did not adjust 
any parameter which was already fixed in the fits to $\pi N$ elastic scattering.  
Some of our results for total cross sections ($\sigma$), 
differential cross sections ($d\sigma/d\Omega)$, and photon 
asymmetry ($\Sigma$) are shown in Fig.~\ref{fig:gnpin}. 

The $Q^2$-dependence of the $\Gamma_{N^*,\gamma N}$ vertex functions were then
determined~\cite{jklmss09} by fitting the $p(e,e'\pi^0)p$ and $p(e,e'\pi^+)n$ data 
up to $W = 1.6$ GeV and $Q^2 = 1.5 $ (GeV/c)$^2$. 

\subsection{Results for two-pions production reactions}

\begin{figure}[t]
\includegraphics[clip,height=.35\textheight]{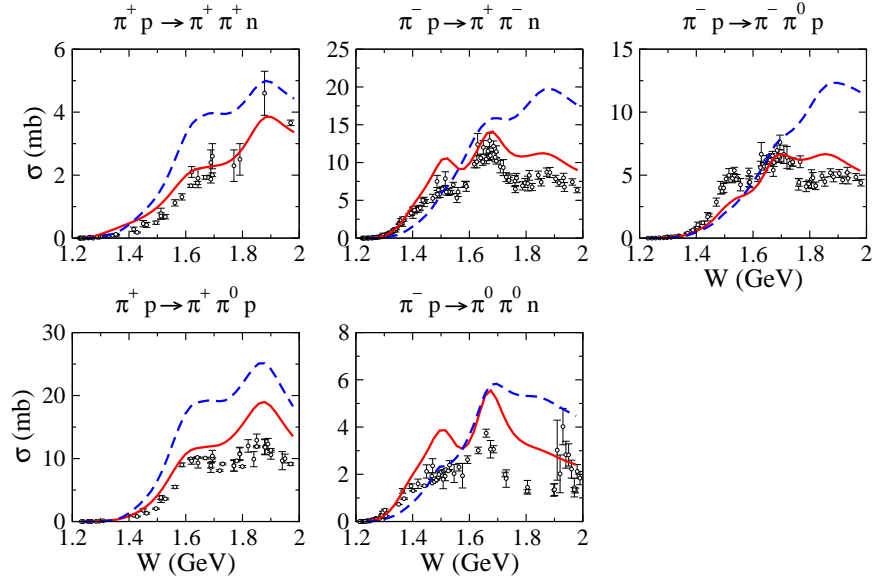}
\caption{The predicted~\cite{kjlms09} total cross sections of the
$\pi N \rightarrow \pi\pi N$ are compared with the data.
The dashed-curves are obtained when the coupled-channel effects are turned
off within the DCC model of Ref.\cite{msl07}.}
\label{fig:pnppn}
\end{figure}

The model constructed from fitting the data on single pion production reactions
was then tested by examining the extent to which the $\pi N \rightarrow \pi\pi N$
and $\gamma N \rightarrow \pi\pi N$ data can be described.
It was found~\cite{kjlms09,kjlms09b} that the predicted total cross sections
are in excellent agreement with the data in the near threshold 
region $W \leq 1.4 $ GeV. 
In the higher $W$ region, the predicted $\pi N \rightarrow \pi\pi N$ cross sections
can describe to a very large extent the available data, as shown in Fig.~\ref{fig:pnppn}. 
Here the important role of the coupled-channel effects were also demonstrated.
However, the predicted $\gamma p \rightarrow \pi^+\pi^- p, \pi^0\pi^0p$ cross sections
were a factor of about 2 larger than the data while the shapes of
two-particles invariant mass distributions could be described very well.

\subsection{Resonance Extractions}

We follow  earlier work, as reviewed and explained in Refs.~\cite{ssl09,ssl10}, 
to define that the resonances are the eigenstates of 
the Hamiltonian with only outgoing waves of their decay channels.
One can then show that the nucleon resonance positions are
the poles $M_R$ of meson-baryon scattering amplitudes calculated
from Eqs.~(\ref{eq:cct}) and~(\ref{eq:ccv}) on 
the unphysical sheets of complex-$E$ Riemann surface.
The coupling of meson-baryon states with the resonances can be
determined by the residues $R_{N^*,MB}$ at the pole positions.
Our procedures for determining $M_R$ and $R_{N^*,MB}$ and the results
were presented in Refs.~\cite{ssl09,ssl10,sjklms10,knls10}.

\begin{figure}[t]
\includegraphics[clip,height=.2\textheight]{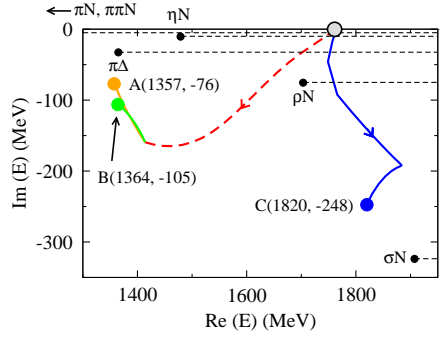}
\caption{The trajectories of the evolution of
three nucleon resonances in $P_{11}$ from the same bare $N^*$ state.
The results are from Ref.~\cite{sjklms10}.}
\label{fig:p11-traj}
\end{figure}

With our analytic continuation method~\cite{ssl09,ssl10},
we were able to analyze the dynamical origins of the nucleon resonances
extracted from the considered DCC model of Ref.\cite{msl07}.
This was done by examining how the resonance positions move as the
coupled-channels effects are gradually turned off. 
As illustrated in Fig.~\ref{fig:p11-traj} for the $P_{11}$ states,
this exercise revealed that the two poles in the Roper region and the next higher
pole are associated with the same bare state.

\begin{figure}[t]
\includegraphics[clip,height=.2\textheight]{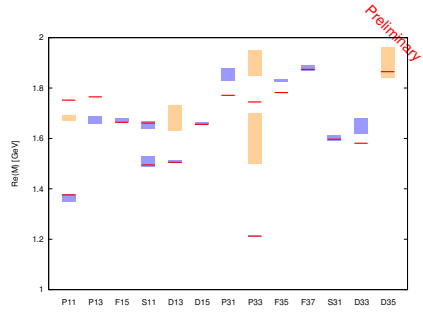}
\caption{Preliminary results (red bars) of the determined $N^*$ spectrum
are compared with 4-star (blue bands) and 3-star (brown bands) states
listed by the Particle Data Group.}
\label{fig:spectrum-1}
\end{figure}

\section{Prospect}

During the developing stage of DCC analysis by Collaboration$@$EBAC
 in 2006-2010,  
the DCC model parameters were determined by analyzing 
separately the following data: $\pi N\rightarrow \pi N$~\cite{jlms07}, 
$\gamma N \rightarrow \pi N$~\cite{jlmss08}, $N(e,e'\pi)N$~\cite{jklmss09},
$\pi N\rightarrow \pi\pi N$~\cite{kjlms09}, 
and $\gamma N \rightarrow \pi\pi N$~\cite{kjlms09b}. 
The very extensive data on $K\Lambda$ and $K\Sigma$ production were not included in the analysis.
To have a high precision extraction of nucleon resonances, it is necessary to
perform a \emph{combined} simultaneous coupled-channels analysis 
of all meson production reactions.

We started an 8-channels combined analysis of \emph{world} data of
$\pi N, \gamma N \rightarrow \pi N, \eta N, K\Lambda, K\Sigma$ 
since the summer of 2010. 
Only the preliminary results had been obtained by March 31, 2012 when
the Collaboration$@$EBAC was dissolved. In Fig.\ref{fig:spectrum-1}, we show
the determined excited nucleon spectrum from this combined analysis.
An ANL-Osaka collaboration  has since taken over this task
and  we expect to complete the analysis by the end of 2012.
The ANL-Osaka analysis will then proceed to
 extracting $\gamma N \rightarrow N^*$ form factors
up to sufficiently high $Q^2$ from  the $new$ JLab data on meson
electroproduction data.
In addition, we will explore the interpretations of the extracted resonance parameters 
in terms of available hadron models,
such as the Dyson-Schwinger-Equation model and constituent quark model, 
and
Lattice QCD.
This last step is needed to complete the DCC project with conclusive results, 
as discussed in Refs.\cite{msl07,sl96,sjklms10,roberts}.

\begin{theacknowledgments}
The author thanks  B.~Juli\'a-D\'iaz, H. Kamano,
 A.~Matsuyama, S.~X.~Nakamura, T.~Sato,
 and N.~Suzuki
for their collaborations at EBAC, and
 would also like to thank A.~W.~Thomas for his strong support 
and his many constructive discussions.
This work is supported by the U.S. Department of Energy, Office of Nuclear Physics Division,
under Contract No. DE-AC02-06CH11357,
and Contract No. DE-AC05-06OR23177 under which Jefferson Science Associates operates the Jefferson Lab.
This research used resources of the National Energy Research Scientific Computing Center, 
which is supported by the Office of Science of the U.S. Department of Energy 
under Contract No. DE-AC02-05CH11231, resources provided on ``Fusion,'' 
a 320-node computing cluster operated by the Laboratory Computing Resource Center 
at Argonne National Laboratory, and resources of Barcelona Sucpercomputing Center (BSC/CNS).
\end{theacknowledgments}

\bibliographystyle{aipproc}

\end{document}